\documentclass[aps,prb,showpacs,floatfix,twocolumn,byrevtex]{revtex4}
\usepackage{amsfonts}
\usepackage{amssymb}
\usepackage{bm}
\usepackage{dcolumn}
\usepackage{graphicx}

\begin{document}

\preprint{Draft, not for distribution}
%
%
%
\title{Scaling of the superfluid density in severely underdoped
  \boldmath YBa$_2$Cu$_3$O$_{6+y}$ \unboldmath}
\author{C. C. Homes}
\email{homes@bnl.gov}
\affiliation{Department of Condensed Matter Physics and Materials Science,
  Brookhaven National Laboratory, Upton, New York 11973, USA}%
%
%
%


%
%
\begin{abstract}
%
%

Recent measurements on extremely-underdoped YBa$_2$Cu$_3$O$_{6+y}$ [Phys. Rev.
Lett. {\bf 99}, 237003 (2007)] have allowed the critical temperature $T_c$, superfluid
density $\rho_{s0}\equiv\rho_s(T\!\ll\!T_c)$ and dc conductivity $\sigma_{dc}(T\gtrsim
T_c)$ to be determined for a series of electronic dopings for $T_c \simeq 3 - 17$~K.
The general scaling relation $\rho_{s0}/8 \simeq 4.4\, \sigma_{dc}T_c$ is observed,
extending the validity of both the {\em ab}-plane and {\em c}-axis scaling an order
of magnitude and creating a region of overlap for the first time.  This suggests that
severely underdoped materials may constitute a Josephson phase;  as the electronic
doping is increased a more uniform superconducting state emerges.

\end{abstract}
%
%
%
%
%
\maketitle

%
%
%
Since the discovery of superconductivity at elevated temperatures in the cuprate materials,
there has been a concerted effort to identify empirical scaling relations in the hope of
providing insights to the mechanism for superconductivity in these materials.  While the
nature of the superconductivity remains elusive, it is now generally accepted
that in the optimally-doped compounds the superconducting energy gap has a {\em d}-wave
symmetry.  Much of the current research in the cuprate materials focuses on the underdoped
compounds where the development of a pseudogap\cite{timusk} in the antinodal region in the
normal state at a characteristic temperature $T^\ast$ leads to the formation of Fermi arcs
(or pockets) centered around the nodal regions.\cite{kanigel06,yang08}  There is
currently a considerable amount of debate as to whether or not the pseudogap represents
preformed pairs that simply lack the coherence required for superconductivity,\cite{yang08}
or if superconductivity originates in the Fermi arcs (pockets) and the pseudogap reflects
some alternative ground state that competes with superconductivity.\cite{kohsaka08,ma08,kondo09}
%
%
An important scaling relation in the underdoped cuprates is the Uemura relation,\cite{uemura89,uemura91}
which notes that the superfluid density at low temperature ($\rho_{s0}$) is proportional to
the critical temperature ($T_c$).  The superfluid density is defined here as $\rho_{s0}
\equiv 1/\lambda_0^2$, where $\lambda_0 = \lambda(T\ll T_c)$ is the effective penetration
depth.  Alternatively, $\rho_{s0}$ is also referred to as the superfluid stiffness where
$\rho_{s0}=4\pi n_{s0} e^2/m^\ast c^2$, where $n_{s0}$ is the density of the superconducting
carriers and $m^\ast$ is an effective mass.  The Uemura relation works well over much
of the underdoped region, but does not apply in optimal and overdoped materials.\cite{tallon03}
In addition, as the phase diagram for YBa$_2$Cu$_3$O$_{6+y}$ has been extended to the severely
underdoped regime,\cite{liang02} the scaling of $\rho_{s0}$ is observed to change from a
linear to power-law relation.\cite{zuev04,liang05,hetel07,broun07}

%
%
In this work we demonstrate that the severely-underdoped data for YBa$_2$Cu$_3$O$_{6+y}$
is in fact consistent with a more general scaling relation $\rho_{s0}/8 \simeq 4.4\,\sigma_{dc}
\,T_c$ where $\sigma_{dc}$ is the dc conductivity measured at $T \gtrsim T_c$ (note that
in this representation both $\sigma_{dc}$ and $T_c$ are shown in units of cm$^{-1}$, so
that the constant is dimensionless and $\rho_{s0}$ has the units of
cm$^{-2}$).\cite{homes04b,homes05a,homes05b}  In the cuprates this scaling relation is
valid for the copper-oxygen planes, as well as along the poorly-conducting {\em c} axis.
The underdoped YBa$_2$Cu$_3$O$_{6+y}$ data extends the validity of both the {\em ab}-plane
and {\em c}-axis scaling by almost an order of magnitude, and also provides a previously
unavailable region of overlap between the {\em a-b} planes and the {\em c} axis.
The possibility of a continuous evolution of the in-plane behavior from a Josephson phase in
the severely underdoped materials to a more uniform superconducting state in systems with
higher electronic dopings is considered.

%
%
In general, the study of the extremely-underdoped region of the phase diagram for the cuprate
materials has been complicated by the cation disorder accompanied by broad transition widths.
An advantage of YBa$_2$Cu$_3$O$_{6+y}$ is that the hole doping can be tuned in a reversible
way by controlling the amount of oxygen in the copper-oxygen chains.  At room temperature, the
dopant oxygens in the chains are mobile and gradually order into chain structures, removing
electrons from the copper-oxygen planes and increasing the hole doping and $T_c$.  Recent
advances in the synthesis of YBa$_2$Cu$_3$O$_{6.333}$ (Ref.~\onlinecite{liang06}) allows the
electronic doping in the copper-oxygen planes to be tuned continuously in a single sample
with no change in the cation disorder and relatively sharp superconducting
transitions.\cite{hosseini04}  Through annealing the chain order can be controlled and $T_c$ may
be increased from $\simeq 3$~K to as high as 17~K and then relaxed back to $\simeq 3$~K; this
is a remarkable achievement in a material with a maximum $T_c \simeq 93\,$K.  In between
periods of annealing, microwave surface impedance techniques have been employed to
determine $T_c$ and the {\em ab}-plane values of $\rho_{s0}$ and $\sigma_{dc}(T\gtrsim T_c)$
for a series of 39 dopings.\cite{broun07}  The {\em c} axis properties have been determined
in a separate experiment for a series of 13 dopings.\cite{huttema}

In the copper-oxygen planes, the maximum electronic doping yields a $T_c = 17.4$~K,
$\sigma_{dc} = 5400$~$\Omega^{-1}{\rm cm}^{-1}$ and $\lambda_0 = 3870$~\AA ; for the
minimum electronic doping $T_c \simeq 3$~K while the dc conductivity is roughly half
of its previous value $\sigma_{dc} = 3020$~$\Omega^{-1}{\rm cm}^{-1}$ and the penetration
depth has increased dramatically to $\lambda_0 \simeq 0.24\,\mu$m.  When $\sigma_{dc}$
is discussed in terms of a sheet resistance $R_{\square} = \rho_{dc}/d$ (where the
interbilayer separation is $d = 11.8$~\AA ) then for the maximum electronic doping
$R_{\square} = 3.2$~k$\Omega$ (per sheet), while the minimum electronic doping yields
$R_{\square} = 5.6$~k$\Omega$, which is remarkably close to the threshold
for the superconductor-insulator transition\cite{strongin70} observed to occur close
to $R_{\square} = h/4e^2 \simeq 6.9$~k$\Omega$ .
The underdoped materials are extremely anisotropic, with $(\lambda_{0,c}/\lambda_{0,ab})^2
\simeq 10^4$ over much of the doping range.  Along the poorly-conducting {\em c} axis,
for the maximum electronic doping  $T_c = 16.9$~K, $\sigma_{dc} = 0.67$~$\Omega^{-1}{\rm cm}^{-1}$
and $\lambda_0 = 40$~$\mu$m, while for the minimum $T_c = 3.9$~K, $\sigma_{dc} =
0.52$~$\Omega^{-1}{\rm cm}^{-1}$  and $\lambda_0 = 137$~$\mu$m.
%

%
%
\begin{figure}[t]
%
%
\vspace*{-0.5cm}%
\centerline{\includegraphics[width=3.6in]{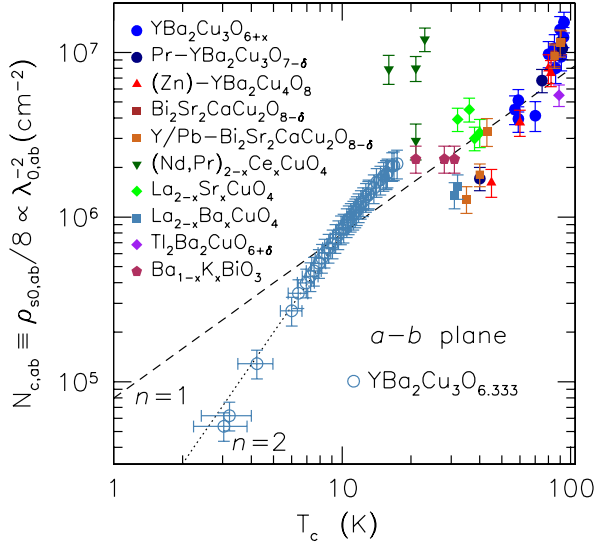}}%
\vspace*{-1.3cm}%
\caption{(Color online)  The log-log plot of the spectral weight of the
superfluid density $N_c \equiv \rho_{s0}/8$ vs $T_c$ for the {\em a-b}
planes for a variety electron- and hole-doped cuprates, as well
as the microwave data for a single crystal of severely underdoped
YBa$_2$Cu$_3$O$_{6.333}$ for 39 different electronic dopings,
yielding a range of $T_c \simeq 3 - 17$~K.  The functional form
$\rho_{s0} \propto T_c^n$ is considered for $n=1$ (dashed line) and
$n=2$ (dotted line).}%
\label{fig:ybco}
\end{figure}

%
%
The optically-determined values of $\rho_{s0}/8$ in the {\em a-b} planes are shown
as a function of $T_c$ in Fig.~\ref{fig:ybco} for a variety of single-layer and double-layer
cuprates.\cite{homes05a}  The quantity $\rho_{s0}/8$ is also referred to as the spectral weight
of the condensate, $N_c$.  The spectral weight is defined as
$$N(\omega_c,T) = \int_{0^+}^{\omega_c} \sigma_1(\omega,T) d\omega,$$
which is simply the area under the conductivity curve over a given interval.  The spectral
weight of the condensate is $N_c = N(\omega,T\simeq T_c) - N(\omega, T\ll T_c)$, where the
cutoff-frequency is chosen so that $N_c$ has converged.  The superfluid density is
precisely $\rho_{s0} \equiv 8N_c$; this transfer of spectral weight is also known as
the Ferrell-Glover-Tinkham sum rule.\cite{ferrell58}
While some of the optical data falls close to the $\rho_{s0} \propto T_c$ (dashed) line in
Fig.~\ref{fig:ybco}, there is a great deal of scatter.\cite{homes05a}  In the optimal and
overdoped materials, there is a clear departure from the linear relation, as has been
previously noted in other works.\cite{tallon03}  The in-plane values of $\rho_{s0}/8$ determined
by microwave techniques for the severely underdoped YBa$_2$Cu$_3$O$_{6.333}$ material are shown
and display a clear $\rho_{s0} \propto T_c^2$ behavior.\cite{zuev04,liang05,hetel07,broun07}
As Fig.~\ref{fig:ybco} illustrates, there is no simple scaling relation between $\rho_{s0}$ and
$T_c$ that is valid over the entire range of electronic dopings.

%
%
\begin{figure}[b]
%
%
\centerline{\includegraphics[width=3.6in]{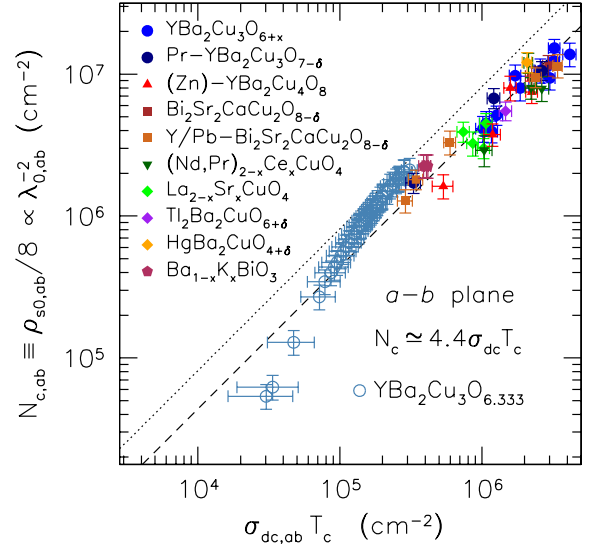}}%
\vspace*{-1.0cm}%
\caption{(Color online) The log-log plot of the spectral weight of the
superfluid density $N_c \equiv \rho_{s0}/8$ vs $\sigma_{dc}\,T_c$ for the
{\em a-b} planes axis for a variety electron- and hole-doped cuprates,
including the severely-underdoped YBa$_2$Cu$_3$O$_{6.333}$ material.
The dotted line is the expected result for a BCS superconductor in the
weak-coupling limit where the normal-state scattering rate is much larger
than the isotropic gap ($N_c = \rho_{s0}/8 \simeq 8.1\,\sigma_{dc}T_c$
\cite{homes05a}); the dashed line is the observed scaling
($N_c = \rho_{s0}/8 \simeq 4.4\,\sigma_{dc}T_c$). }%
\label{fig:abplane}
\end{figure}

The values for $\rho_{s0}/8$ in the copper-oxygen planes in Fig.~\ref{fig:ybco} have been
replotted as a function of $\sigma_{dc}\,T_c$ in Fig.~\ref{fig:abplane} [in the optical
and microwave measurements, $\sigma_{dc}\equiv \sigma_1(\omega \rightarrow 0)$ at
$T\gtrsim T_c$].  The dashed line in Fig.~\ref{fig:abplane} is the best fit to the data,
$N_c = \rho_{s0}/8 \simeq 4.4\,\sigma_{dc}T_c$, while the dotted line is the calculated
result $N_c = \rho_{s0}/8 \simeq 8.1\,\sigma_{dc}T_c$ for a BCS superconductor where
the normal-state scattering rate $1/\tau$ (taken at $T\gtrsim T_c$) is greater than
the isotropic gap $2\Delta$ in the weak-coupling limit;\cite{homes05a} this is
equivalent to the so-called ``dirty-limit'' condition that $l \lesssim \xi_0$,
where $l=v_F\tau$ is the mean-free path, and $\xi_0=\hbar \,v_F/\pi\Delta_0$ is
the coherence length ($v_F$ is the Fermi velocity).
The importance of this result lies in the observation that for $T\ll\,T_c$ there is always
a dramatic suppression of the low frequency optical conductivity.\cite{basov05}
This ``missing area'' under the conductivity curve upon entering the superconducting
state is a consequence of the transfer of normal-state spectral weight into the condensate;
the fact that this transfer may be observed at all is due to the self-consistent condition
that $1/\tau \gtrsim 2\Delta_0$ (Ref.~\onlinecite{kamaras90}).
From this argument it is a relatively straightforward matter to construct a
geometric scaling relation based on the transfer of spectral weight that
yields the observed $\rho_{s0}/8 \propto \sigma_{dc}T_c$ relation.\cite{homes05a}
This is an important result in that it allows statements to be made about the
nature of the superconductivity, and negates arguments based on the assumption
that $1/\tau \ll 2\Delta$ ($2\Delta_0$ in a {\em d}-wave system) that assert
that the scaling relation only contains information about the normal state.
An interesting trend in the scaling of the microwave data is an increase in the
slope at lower dopings; however, if the last three points are neglected (dopings
for which $T_c \lesssim 6$~K), then this trend becomes less noticeable.  Overall,
the new in-plane microwave data for YBa$_2$Cu$_3$O$_{6.333}$ agrees quite well with the
optically-determined results for other underdoped materials and provides a region of
substantial overlap, illustrating that microwave techniques provide a complementary method
for the determination of $\rho_{s0}$ and $\sigma_{dc}$ (this is especially useful when
$\rho_{s0}$ may be too small to be determined accurately using optical techniques).
In addition, the new data extends the validity of the scaling within the copper-oxygen
planes by nearly an order of magnitude.  In comparison, the Uemura scaling shown in
Fig.~\ref{fig:ybco} constitutes is only about half a cycle in Fig.~\ref{fig:abplane}.

\begin{figure}[t]
%
%
\vspace*{-0.5cm}%
\centerline{\includegraphics[width=3.6in]{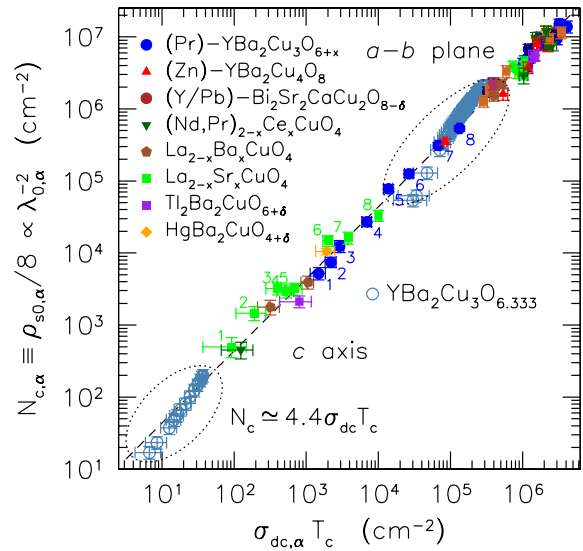}}%
\vspace*{-1.0cm}%
\caption{(Color online) The log-log plot of the spectral weight of the
superfluid density $N_{c,\alpha} \equiv \rho_{s0,\alpha}/8$ vs
$\sigma_{dc,\alpha}\,T_c$, where $\alpha$ denotes either the {\em a-b} plane or
the {\em c}-axis direction, for a variety of electron- and hole-doped
cuprates.  The dashed line corresponds to $\rho_{s0,\alpha}/8 \simeq 4.4\,
\sigma_{dc,\alpha}\,T_c$.  The {\em a-b} plane data for severely-underdoped
YBa$_2$Cu$_3$O$_{6.333}$ sample merges smoothly with the existing optical data
and provides a substantial region of overlap with the {\em c} axis results.  In
addition the scaling along the {\em c} axis has been extended by nearly an
order of magnitude.  The numbers next to the {\em c}-axis data for
La$_{2-x}$Sr$_{x}$CuO$_4$ and YBa$_2$Cu$_3$O$_{6+y}$ correspond to different
chemical compositions (electronic doping) \cite{homes04b,homes05a,homes05b}.
}%
\label{fig:total}
\end{figure}

%
%
The values for $\rho_{s0,\alpha}$ are plotted against $\sigma_{dc,\alpha}T_c$ in
Fig.~\ref{fig:total}, where $\alpha$ denotes either the {\em a-b} plane or
{\em c} axis direction (the regions encompassed by the new microwave data are
denoted by the enclosed regions).  As previously shown in Fig.~\ref{fig:abplane},
the new results merge smoothly with existing in-plane data and extends the results
so that for the first time there is now a significant overlap with the {\em c} axis data.
Furthermore, the new results for YBa$_2$Cu$_3$O$_{6.333}$ along the {\em c} axis
extend the overall scaling by a further order of magnitude so that the general
scaling relation is now valid over nearly six orders of magnitude.  As previously observed
for the in-plane results, the last two points of the most severely underdoped data
($T_c \lesssim 6$~K) along the {\em c} axis may also fall slightly below the
scaling line.
%
%
It has been previously noted that the scaling observed in the {\em a-b} planes
and along the {\em c} axis is the same; $N_c \equiv \rho_{s0}/8 \simeq 4.4\,\sigma_{dc}
T_c$ (Ref.~\onlinecite{homes04b}).  This is surprising given that the normal-state transport
in the planes is coherent and the superconductivity is accompanied by the formation of an
energy gap with {\em d}-wave symmetry,\cite{hardy93,harlingen95} while the normal-state
transport perpendicular to the copper-oxygen layers along the {\em c} axis is governed
by hopping and the superconductivity is due to the Josephson effect.\cite{basov94a,%
shibauchi94}  It has been previously demonstrated\cite{homes05a,homes05b} that a BCS
superconductor in the weak-coupling limit yields the same linear scaling relation
for both the dirty limit ({\em a-b} planes) as well as for tunneling between the
planes due to the Josephson effect ({\em c} axis); $\rho_{s0}/8 \simeq 8.1\,\sigma_{dc}T_c$
(note that the calculated constant is somewhat larger than the experimentally-observed
value).
Until this point, it was never possible to reduce the electronic doping in the
copper-oxygen planes to the extent that there was any region of overlap between the
{\em ab}-planes an the {\em c} axis data. However, as Fig.~\ref{fig:total} demonstrates,
the electronic doping in YBa$_2$Cu$_3$O$_{6.333}$ has been severely reduced, extending
the in-plane results by nearly an order of magnitude and creating a substantial region of
overlap with the {\em c}-axis data.  This suggests that in the most severely underdoped
case the superconductivity in the copper-oxygen planes may be inhomogeneous and dominated
by Josephson effects (a ``Josephson phase''), evolving with increasing doping into a more
uniform superconductor.

Such a progression is not difficult to envision.  It is generally accepted that the
underdoped cuprates are electronically inhomogeneous,\cite{carr89,lang02, mcelroy05a,dagotto05}
and some materials even display static charge- and spin-stripe order in which the material
is segregated into hole-rich and hole-poor regions;\cite{zaanen89,tranquada95} such a
scenario has also been described in terms of frustrated phase separation.\cite{emery93}
We speculate that the severely-underdoped materials are electronically segregated into
superconducting hole-rich regions, and hole-poor regions that form a poorly-conducting
barrier region.  If the ``granularity'' of such a system is fine enough, then the
superconducting regions will be linked through the Josephson effect,\cite{simanek} in
essence forming a Josephson phase,\cite{imry08} and in fact it has recently been
demonstrated that the scaling relation observed here can be derived for a two-dimensional
Josephson array.\cite{imry}  Within this framework the doping level may in principle
be reduced to such an extent that Josephson coupling between the superconducting regions
is no longer possible.  In practice, this would correspond to doping levels below
$\sim 0.05$ holes per planar copper atom.  At this critical doping one would also expect
that the sheet resistance would approach $h/4e^2 \simeq  6.9$~k$\Omega$,
close to the value of $R_\square \simeq 5.6$~k$\Omega$ observed at the lowest doping.
However, as the electronic doping is increased the material becomes more homogeneous,
the size of the superconducting regions increases and the ``granularity'' is reduced,
allowing the system to revert to a more conventional behavior.  It should be noted that
in this ``large-grain'' picture, the Uemura relation is expected to be recovered,\cite{imry08}
suggesting that this is a reasonable description of the moderately-underdoped region.

%
%

In summary the superfluid density in severely underdoped YBa$_2$Cu$_3$O$_{6.333}$ is found
to follow the general scaling relation $\rho_{s0}/8 \simeq 4.4\, \sigma_{dc}T_c$, extending
the validity of the scaling in the copper-oxygen planes and the {\em c} axis by nearly an
order of magnitude, and providing for the first time a region of substantial overlap
between the {\em a-b} plane and {\em c} axis data.  We speculate that in-plane response
in the severely underdoped region is electronically inhomogeneous and that the
superconductivity in this region may constitute a Josephson phase.  However, as the doping
is increased a more homogeneous electronic state emerges and the superconductivity
becomes more uniform.

%
%
The author is deeply indebted to D. M. Broun, W. A. Huttema, P. J. Turner
for providing their their microwave results on extremely underdoped
YBa$_2$Cu$_3$O$_{6.333}$, and to Ruixing Liang, W. N. Hardy and D. A. Bonn
regarding aspects of materials synthesis.  The author would also like to
acknowledge useful discussions with A. Akrap, A. V. Chubukov, G. L. Carr,
Y. Imry, T. R. Lemberger, P. Littlewood, J. Rameau, M. Strongin,
D. B. Tanner and A. M. Tsvelik.
Work in Canada was supported by the Natural Sciences and Engineering Research
Council of Canada and the Canadian Institute for Advanced Research.  Work at
BNL is supported by the Office of Science, U.S. Department of Energy (DOE)
under Contract No. DE-AC02-98CH10886.

%
%
%
\bibliography{scaling}

\begin{thebibliography}{10}%
\makeatletter
\providecommand \@ifxundefined [1]{%
 \ifx #1\undefined \expandafter \@firstoftwo
 \else \expandafter \@secondoftwo
\fi
}%
\providecommand \@ifnum [1]{%
 \ifnum #1\expandafter \@firstoftwo
 \else \expandafter \@secondoftwo
\fi
}%
\providecommand \enquote [1]{``#1''}%
\providecommand \bibnamefont  [1]{#1}%
\providecommand \bibfnamefont [1]{#1}%
\providecommand \citenamefont [1]{#1}%
\providecommand\href[0]{\@sanitize\@href}%
\providecommand\@href[1]{\endgroup\@@startlink{#1}\endgroup\@@href}%
\providecommand\@@href[1]{#1\@@endlink}%
\providecommand \@sanitize [0]{\begingroup\catcode`\&12\catcode`\#12\relax}%
\@ifxundefined \pdfoutput {\@firstoftwo}{%
 \@ifnum{\z@=\pdfoutput}{\@firstoftwo}{\@secondoftwo}%
}{%
 \providecommand\@@startlink[1]{\leavevmode\special{html:<a href="#1">}}%
 \providecommand\@@endlink[0]{\special{html:</a>}}%
}{%
 \providecommand\@@startlink[1]{%
  \leavevmode
  \pdfstartlink
   attr{/Border[0 0 1 ]/H/I/C[0 1 1]}%
   user{/Subtype/Link/A<</Type/Action/S/URI/URI(#1)>>}%
  \relax
 }%
 \providecommand\@@endlink[0]{\pdfendlink}%
}%
\providecommand \url  [0]{\begingroup\@sanitize \@url }%
\providecommand \@url [1]{\endgroup\@href {#1}{\urlprefix}}%
\providecommand \urlprefix [0]{URL }%
\providecommand \Eprint[0]{\href }%
\@ifxundefined \urlstyle {%
  \providecommand \doi [1]{doi:\discretionary{}{}{}#1}%
}{%
  \providecommand \doi [0]{doi:\discretionary{}{}{}\begingroup
  \urlstyle{rm}\Url }%
}%
\providecommand \doibase [0]{http://dx.doi.org/}%
\providecommand \Doi[1]{\href{\doibase#1}}%
\providecommand \bibAnnote [3]{%
  \BibitemShut{#1}%
  \begin{quotation}\noindent
    \textsc{Key:}\ #2\\\textsc{Annotation:}\ #3%
  \end{quotation}%
}%
\providecommand \bibAnnoteFile [2]{%
  \IfFileExists{#2}{\bibAnnote {#1} {#2} {\input{#2}}}{}%
}%
\providecommand \typeout [0]{\immediate \write \m@ne }%
\providecommand \selectlanguage [0]{\@gobble}%
\providecommand \bibinfo [0]{\@secondoftwo}%
\providecommand \bibfield [0]{\@secondoftwo}%
\providecommand \translation [1]{[#1]}%
\providecommand \BibitemOpen[0]{}%
\providecommand \bibitemStop [0]{}%
\providecommand \bibitemNoStop [0]{.\EOS\space}%
\providecommand \EOS [0]{\spacefactor3000\relax}%
\providecommand \BibitemShut [1]{\csname bibitem#1\endcsname}%
\bibitem{timusk}%
  \BibitemOpen
  \bibfield{author}{%
  \bibinfo {author} {\bibfnamefont{T.}~\bibnamefont{Timusk}}\ and\ \bibinfo
  {author} {\bibfnamefont{B.}~\bibnamefont{Statt}},\ }%
  \bibfield{journal}{%
  \bibinfo {journal} {Rep. Prog. Phys.}\ }%
  \textbf{\bibinfo {volume} {62}},\ \bibinfo {pages} {61} (\bibinfo {year}
  {1999})%
  \bibAnnoteFile{NoStop}{timusk}%
\bibitem{kanigel06}%
  \BibitemOpen
  \bibfield{author}{%
  \bibinfo {author} {\bibfnamefont{A.}~\bibnamefont{Kanigel}}, \bibinfo
  {author} {\bibfnamefont{M.~R.}\ \bibnamefont{Norman}}, \bibinfo {author}
  {\bibfnamefont{M.}~\bibnamefont{Randeria}}, \bibinfo {author}
  {\bibfnamefont{U.}~\bibnamefont{Chatterjee}}, \bibinfo {author}
  {\bibfnamefont{S.}~\bibnamefont{Suoma}}, \bibinfo {author}
  {\bibfnamefont{A.}~\bibnamefont{Kaminski}}, \bibinfo {author}
  {\bibfnamefont{H.~M.}\ \bibnamefont{Fretwell}}, \bibinfo {author}
  {\bibfnamefont{S.}~\bibnamefont{Rosenkranz}}, \bibinfo {author}
  {\bibfnamefont{M.}~\bibnamefont{Shi}}, \bibinfo {author}
  {\bibfnamefont{T.}~\bibnamefont{Sato}}, \bibinfo {author}
  {\bibfnamefont{T.}~\bibnamefont{Takahashi}}, \bibinfo {author}
  {\bibfnamefont{Z.~Z.}\ \bibnamefont{Li}}, \bibinfo {author}
  {\bibfnamefont{H.}~\bibnamefont{Raffy}}, \bibinfo {author}
  {\bibfnamefont{K.}~\bibnamefont{Kadowaki}}, \bibinfo {author}
  {\bibfnamefont{D.}~\bibnamefont{Hinks}}, \bibinfo {author}
  {\bibfnamefont{L.}~\bibnamefont{Ozyuzer}},\ and\ \bibinfo {author}
  {\bibfnamefont{J.~C.}\ \bibnamefont{Campuzano}},\ }%
  \bibfield{journal}{%
  \bibinfo {journal} {Nature Phys.}\ }%
  \textbf{\bibinfo {volume} {2}},\ \bibinfo {pages} {447} (\bibinfo {year}
  {2006})%
  \bibAnnoteFile{NoStop}{kanigel06}%
\bibitem{yang08}%
  \BibitemOpen
  \bibfield{author}{%
  \bibinfo {author} {\bibfnamefont{H.-B.}\ \bibnamefont{Yang}}, \bibinfo
  {author} {\bibfnamefont{J.~D.}\ \bibnamefont{Rameau}}, \bibinfo {author}
  {\bibfnamefont{P.~D.}\ \bibnamefont{Johnson}}, \bibinfo {author}
  {\bibfnamefont{T.}~\bibnamefont{Valla}}, \bibinfo {author}
  {\bibfnamefont{A.}~\bibnamefont{Tsvelik}},\ and\ \bibinfo {author}
  {\bibfnamefont{G.~D.}\ \bibnamefont{Gu}},\ }%
  \bibfield{journal}{%
  \bibinfo {journal} {Nature (London)}\ }%
  \textbf{\bibinfo {volume} {456}},\ \bibinfo {pages} {77} (\bibinfo {year}
  {2008})%
  \bibAnnoteFile{NoStop}{yang08}%
\bibitem{kohsaka08}%
  \BibitemOpen
  \bibfield{author}{%
  \bibinfo {author} {\bibfnamefont{Y.}~\bibnamefont{Kohsaka}}, \bibinfo
  {author} {\bibfnamefont{C.}~\bibnamefont{Taylor}}, \bibinfo {author}
  {\bibfnamefont{P.}~\bibnamefont{Wahl}}, \bibinfo {author}
  {\bibfnamefont{A.}~\bibnamefont{Schmidt}}, \bibinfo {author}
  {\bibfnamefont{J.}~\bibnamefont{Lee}}, \bibinfo {author}
  {\bibfnamefont{K.}~\bibnamefont{Fujita}}, \bibinfo {author}
  {\bibfnamefont{J.~W.}\ \bibnamefont{Alldredge}}, \bibinfo {author}
  {\bibfnamefont{K.}~\bibnamefont{McElroy}}, \bibinfo {author}
  {\bibfnamefont{J.}~\bibnamefont{Lee}}, \bibinfo {author}
  {\bibfnamefont{H.}~\bibnamefont{Eisaki}}, \bibinfo {author}
  {\bibfnamefont{S.}~\bibnamefont{Uchida}}, \bibinfo {author}
  {\bibfnamefont{D.-H.}\ \bibnamefont{Lee}},\ and\ \bibinfo {author}
  {\bibfnamefont{J.~C.}\ \bibnamefont{Davis}},\ }%
  \bibfield{journal}{%
  \bibinfo {journal} {Nature (London)}\ }%
  \textbf{\bibinfo {volume} {454}},\ \bibinfo {pages} {1072} (\bibinfo {year}
  {2008})%
  \bibAnnoteFile{NoStop}{kohsaka08}%
\bibitem{ma08}%
  \BibitemOpen
  \bibfield{author}{%
  \bibinfo {author} {\bibfnamefont{J.-H.}\ \bibnamefont{Ma}}, \bibinfo {author}
  {\bibfnamefont{Z.-H.}\ \bibnamefont{Pan}}, \bibinfo {author}
  {\bibfnamefont{F.~C.}\ \bibnamefont{Niestemski}}, \bibinfo {author}
  {\bibfnamefont{M.}~\bibnamefont{Neupane}}, \bibinfo {author}
  {\bibfnamefont{Y.-M.}\ \bibnamefont{Xu}}, \bibinfo {author}
  {\bibfnamefont{P.}~\bibnamefont{Richard}}, \bibinfo {author}
  {\bibfnamefont{K.}~\bibnamefont{Nakayama}}, \bibinfo {author}
  {\bibfnamefont{T.}~\bibnamefont{Sato}}, \bibinfo {author}
  {\bibfnamefont{T.}~\bibnamefont{Takahashi}}, \bibinfo {author}
  {\bibfnamefont{H.-Q.}\ \bibnamefont{Luo}}, \bibinfo {author}
  {\bibfnamefont{L.}~\bibnamefont{Fang}}, \bibinfo {author}
  {\bibfnamefont{H.-H.}\ \bibnamefont{Wen}}, \bibinfo {author}
  {\bibfnamefont{Z.}~\bibnamefont{Wang}}, \bibinfo {author}
  {\bibfnamefont{H.}~\bibnamefont{Ding}},\ and\ \bibinfo {author}
  {\bibfnamefont{V.}~\bibnamefont{Madhavan}},\ }%
  \bibfield{journal}{%
  \bibinfo {journal} {Phys. Rev. Lett.}\ }%
  \textbf{\bibinfo {volume} {101}},\ \bibinfo {eid} {207002} (\bibinfo {year}
  {2008})%
  \bibAnnoteFile{NoStop}{ma08}%
\bibitem{kondo09}%
  \BibitemOpen
  \bibfield{author}{%
  \bibinfo {author} {\bibfnamefont{T.}~\bibnamefont{Kondo}}, \bibinfo {author}
  {\bibfnamefont{R.}~\bibnamefont{Khasanov}}, \bibinfo {author}
  {\bibfnamefont{T.}~\bibnamefont{Takeuchi}}, \bibinfo {author}
  {\bibfnamefont{J.}~\bibnamefont{Schmalian}},\ and\ \bibinfo {author}
  {\bibfnamefont{A.}~\bibnamefont{Kaminski}},\ }%
  \bibfield{journal}{%
  \bibinfo {journal} {Nature (London)}\ }%
  \textbf{\bibinfo {volume} {457}},\ \bibinfo {pages} {296} (\bibinfo {year}
  {2009})%
  \bibAnnoteFile{NoStop}{kondo09}%
\bibitem{uemura89}%
  \BibitemOpen
  \bibfield{author}{%
  \bibinfo {author} {\bibfnamefont{Y.~J.}\ \bibnamefont{Uemura}}, \bibinfo
  {author} {\bibfnamefont{G.~M.}\ \bibnamefont{Luke}}, \bibinfo {author}
  {\bibfnamefont{B.~J.}\ \bibnamefont{Sternlieb}}, \bibinfo {author}
  {\bibfnamefont{J.~H.}\ \bibnamefont{Brewer}}, \bibinfo {author}
  {\bibfnamefont{J.~F.}\ \bibnamefont{Carolan}}, \bibinfo {author}
  {\bibfnamefont{W.~N.}\ \bibnamefont{Hardy}}, \bibinfo {author}
  {\bibfnamefont{R.}~\bibnamefont{Kadono}}, \bibinfo {author}
  {\bibfnamefont{J.~R.}\ \bibnamefont{Kempton}}, \bibinfo {author}
  {\bibfnamefont{R.~F.}\ \bibnamefont{Kiefl}}, \bibinfo {author}
  {\bibfnamefont{S.~R.}\ \bibnamefont{Kreitzman}}, \bibinfo {author}
  {\bibfnamefont{P.}~\bibnamefont{Mulhern}}, \bibinfo {author}
  {\bibfnamefont{T.~M.}\ \bibnamefont{Riseman}}, \bibinfo {author}
  {\bibfnamefont{D.~L.}\ \bibnamefont{Williams}}, \bibinfo {author}
  {\bibfnamefont{B.~X.}\ \bibnamefont{Yang}}, \bibinfo {author}
  {\bibfnamefont{S.}~\bibnamefont{Uchida}}, \bibinfo {author}
  {\bibfnamefont{H.}~\bibnamefont{Takagi}}, \bibinfo {author}
  {\bibfnamefont{J.}~\bibnamefont{Gopalakrishnan}}, \bibinfo {author}
  {\bibfnamefont{A.~W.}\ \bibnamefont{Sleight}}, \bibinfo {author}
  {\bibfnamefont{M.~A.}\ \bibnamefont{Subramanian}}, \bibinfo {author}
  {\bibfnamefont{C.~L.}\ \bibnamefont{Chien}}, \bibinfo {author}
  {\bibfnamefont{M.~Z.}\ \bibnamefont{Cieplak}}, \bibinfo {author}
  {\bibfnamefont{G.}~\bibnamefont{Xiao}}, \bibinfo {author}
  {\bibfnamefont{V.~Y.}\ \bibnamefont{Lee}}, \bibinfo {author}
  {\bibfnamefont{B.~W.}\ \bibnamefont{Statt}}, \bibinfo {author}
  {\bibfnamefont{C.~E.}\ \bibnamefont{Stronach}}, \bibinfo {author}
  {\bibfnamefont{W.~J.}\ \bibnamefont{Kossler}},\ and\ \bibinfo {author}
  {\bibfnamefont{X.~H.}\ \bibnamefont{Yu}},\ }%
  \bibfield{journal}{%
  \bibinfo {journal} {Phys. Rev. Lett.}\ }%
  \textbf{\bibinfo {volume} {62}},\ \bibinfo {pages} {2317} (\bibinfo {year}
  {1989})%
  \bibAnnoteFile{NoStop}{uemura89}%
\bibitem{uemura91}%
  \BibitemOpen
  \bibfield{author}{%
  \bibinfo {author} {\bibfnamefont{Y.~J.}\ \bibnamefont{Uemura}}, \bibinfo
  {author} {\bibfnamefont{L.~P.}\ \bibnamefont{Le}}, \bibinfo {author}
  {\bibfnamefont{G.~M.}\ \bibnamefont{Luke}}, \bibinfo {author}
  {\bibfnamefont{B.~J.}\ \bibnamefont{Sternlieb}}, \bibinfo {author}
  {\bibfnamefont{W.~D.}\ \bibnamefont{Wu}}, \bibinfo {author}
  {\bibfnamefont{J.~H.}\ \bibnamefont{Brewer}}, \bibinfo {author}
  {\bibfnamefont{T.~M.}\ \bibnamefont{Riseman}}, \bibinfo {author}
  {\bibfnamefont{C.~L.}\ \bibnamefont{Seaman}}, \bibinfo {author}
  {\bibfnamefont{M.~B.}\ \bibnamefont{Maple}}, \bibinfo {author}
  {\bibfnamefont{M.}~\bibnamefont{Ishikawa}}, \bibinfo {author}
  {\bibfnamefont{D.~G.}\ \bibnamefont{Hinks}}, \bibinfo {author}
  {\bibfnamefont{J.~D.}\ \bibnamefont{Jorgensen}}, \bibinfo {author}
  {\bibfnamefont{G.}~\bibnamefont{Saito}},\ and\ \bibinfo {author}
  {\bibfnamefont{H.}~\bibnamefont{Yamochi}},\ }%
  \bibfield{journal}{%
  \bibinfo {journal} {Phys. Rev. Lett.}\ }%
  \textbf{\bibinfo {volume} {66}},\ \bibinfo {pages} {2665} (\bibinfo {year}
  {1991})%
  \bibAnnoteFile{NoStop}{uemura91}%
\bibitem{tallon03}%
  \BibitemOpen
  \bibfield{author}{%
  \bibinfo {author} {\bibfnamefont{J.~L.}\ \bibnamefont{Tallon}}, \bibinfo
  {author} {\bibfnamefont{J.~W.}\ \bibnamefont{Loram}}, \bibinfo {author}
  {\bibfnamefont{J.~R.}\ \bibnamefont{Cooper}}, \bibinfo {author}
  {\bibfnamefont{C.}~\bibnamefont{Panagopoulos}},\ and\ \bibinfo {author}
  {\bibfnamefont{C.}~\bibnamefont{Bernhard}},\ }%
  \bibfield{journal}{%
  \bibinfo {journal} {Phys. Rev. B}\ }%
  \textbf{\bibinfo {volume} {68}},\ \bibinfo {pages} {180501(R)} (\bibinfo
  {year} {2003})%
  \bibAnnoteFile{NoStop}{tallon03}%
\bibitem{liang02}%
  \BibitemOpen
  \bibfield{author}{%
  \bibinfo {author} {\bibfnamefont{R.}~\bibnamefont{Liang}},\ }%
  \bibfield{journal}{%
  \bibinfo {journal} {Physica C}\ }%
  \textbf{\bibinfo {volume} {383}},\ \bibinfo {pages} {1} (\bibinfo {year}
  {2002})%
  \bibAnnoteFile{NoStop}{liang02}%
\bibitem{zuev04}%
  \BibitemOpen
  \bibfield{author}{%
  \bibinfo {author} {\bibfnamefont{Y.}~\bibnamefont{Zuev}}, \bibinfo {author}
  {\bibfnamefont{M.~S.}\ \bibnamefont{Kim}},\ and\ \bibinfo {author}
  {\bibfnamefont{T.~R.}\ \bibnamefont{Lemberger}},\ }%
  \bibfield{journal}{%
  \bibinfo {journal} {Phys. Rev. Lett.}\ }%
  \textbf{\bibinfo {volume} {95}},\ \bibinfo {pages} {137002} (\bibinfo {year}
  {2005})%
  \bibAnnoteFile{NoStop}{zuev04}%
\bibitem{liang05}%
  \BibitemOpen
  \bibfield{author}{%
  \bibinfo {author} {\bibfnamefont{R.}~\bibnamefont{Liang}}, \bibinfo {author}
  {\bibfnamefont{D.~A.}\ \bibnamefont{Bonn}}, \bibinfo {author}
  {\bibfnamefont{W.~N.}\ \bibnamefont{Hardy}},\ and\ \bibinfo {author}
  {\bibfnamefont{D.}~\bibnamefont{Broun}},\ }%
  \bibfield{journal}{%
  \bibinfo {journal} {Phys. Rev. Lett.}\ }%
  \textbf{\bibinfo {volume} {94}},\ \bibinfo {pages} {117001} (\bibinfo {year}
  {2005})%
  \bibAnnoteFile{NoStop}{liang05}%
\bibitem{hetel07}%
  \BibitemOpen
  \bibfield{author}{%
  \bibinfo {author} {\bibfnamefont{I.}~\bibnamefont{Hetel}}, \bibinfo {author}
  {\bibfnamefont{T.~R.}\ \bibnamefont{Lemberger}},\ and\ \bibinfo {author}
  {\bibfnamefont{M.}~\bibnamefont{Randeria}},\ }%
  \bibfield{journal}{%
  \bibinfo {journal} {Nature Phys.}\ }%
  \textbf{\bibinfo {volume} {3}},\ \bibinfo {pages} {700} (\bibinfo {year}
  {2007})%
  \bibAnnoteFile{NoStop}{hetel07}%
\bibitem{broun07}%
  \BibitemOpen
  \bibfield{author}{%
  \bibinfo {author} {\bibfnamefont{D.~M.}\ \bibnamefont{Broun}}, \bibinfo
  {author} {\bibfnamefont{W.~A.}\ \bibnamefont{Huttema}}, \bibinfo {author}
  {\bibfnamefont{P.~J.}\ \bibnamefont{Turner}}, \bibinfo {author}
  {\bibfnamefont{S.}~\bibnamefont{\"{O}zcan}}, \bibinfo {author}
  {\bibfnamefont{B.}~\bibnamefont{Morgan}}, \bibinfo {author}
  {\bibfnamefont{R.}~\bibnamefont{Liang}}, \bibinfo {author}
  {\bibfnamefont{W.~N.}\ \bibnamefont{Hardy}},\ and\ \bibinfo {author}
  {\bibfnamefont{D.~A.}\ \bibnamefont{Bonn}},\ }%
  \bibfield{journal}{%
  \bibinfo {journal} {Phys. Rev. Lett.}\ }%
  \textbf{\bibinfo {volume} {99}},\ \bibinfo {eid} {237003} (\bibinfo {year}
  {2007})%
  \bibAnnoteFile{NoStop}{broun07}%
\bibitem{homes04b}%
  \BibitemOpen
  \bibfield{author}{%
  \bibinfo {author} {\bibfnamefont{C.~C.}\ \bibnamefont{Homes}}, \bibinfo
  {author} {\bibfnamefont{S.~V.}\ \bibnamefont{Dordevic}}, \bibinfo {author}
  {\bibfnamefont{M.}~\bibnamefont{Strongin}}, \bibinfo {author}
  {\bibfnamefont{D.~A.}\ \bibnamefont{Bonn}}, \bibinfo {author}
  {\bibfnamefont{R.}~\bibnamefont{Liang}}, \bibinfo {author}
  {\bibfnamefont{W.~N.}\ \bibnamefont{Hardy}}, \bibinfo {author}
  {\bibfnamefont{S.}~\bibnamefont{Komiya}}, \bibinfo {author}
  {\bibfnamefont{Y.}~\bibnamefont{Ando}}, \bibinfo {author}
  {\bibfnamefont{G.}~\bibnamefont{Yu}}, \bibinfo {author}
  {\bibfnamefont{N.}~\bibnamefont{Kaneko}}, \bibinfo {author}
  {\bibfnamefont{X.}~\bibnamefont{Zhao}}, \bibinfo {author}
  {\bibfnamefont{M.}~\bibnamefont{Greven}}, \bibinfo {author}
  {\bibfnamefont{D.~N.}\ \bibnamefont{Basov}},\ and\ \bibinfo {author}
  {\bibfnamefont{T.}~\bibnamefont{Timusk}},\ }%
  \bibfield{journal}{%
  \bibinfo {journal} {Nature (London)}\ }%
  \textbf{\bibinfo {volume} {430}},\ \bibinfo {pages} {539} (\bibinfo {year}
  {2004})%
  \bibAnnoteFile{NoStop}{homes04b}%
\bibitem{homes05a}%
  \BibitemOpen
  \bibfield{author}{%
  \bibinfo {author} {\bibfnamefont{C.~C.}\ \bibnamefont{Homes}}, \bibinfo
  {author} {\bibfnamefont{S.~V.}\ \bibnamefont{Dordevic}}, \bibinfo {author}
  {\bibfnamefont{T.}~\bibnamefont{Valla}},\ and\ \bibinfo {author}
  {\bibfnamefont{M.}~\bibnamefont{Strongin}},\ }%
  \bibfield{journal}{%
  \bibinfo {journal} {Phys. Rev. B}\ }%
  \textbf{\bibinfo {volume} {72}},\ \bibinfo {pages} {134517} (\bibinfo {year}
  {2005})%
  \bibAnnoteFile{NoStop}{homes05a}%
\bibitem{homes05b}%
  \BibitemOpen
  \bibfield{author}{%
  \bibinfo {author} {\bibfnamefont{C.~C.}\ \bibnamefont{Homes}}, \bibinfo
  {author} {\bibfnamefont{S.~V.}\ \bibnamefont{Dordevic}}, \bibinfo {author}
  {\bibfnamefont{D.~A.}\ \bibnamefont{Bonn}}, \bibinfo {author}
  {\bibfnamefont{R.}~\bibnamefont{Liang}}, \bibinfo {author}
  {\bibfnamefont{W.~N.}\ \bibnamefont{Hardy}},\ and\ \bibinfo {author}
  {\bibfnamefont{T.}~\bibnamefont{Timusk}},\ }%
  \bibfield{journal}{%
  \bibinfo {journal} {Phys. Rev. B}\ }%
  \textbf{\bibinfo {volume} {71}},\ \bibinfo {pages} {184515} (\bibinfo {year}
  {2005})%
  \bibAnnoteFile{NoStop}{homes05b}%
\bibitem{liang06}%
  \BibitemOpen
  \bibfield{author}{%
  \bibinfo {author} {\bibfnamefont{R.}~\bibnamefont{Liang}}, \bibinfo {author}
  {\bibfnamefont{D.~A.}\ \bibnamefont{Bonn}},\ and\ \bibinfo {author}
  {\bibfnamefont{W.~N.}\ \bibnamefont{Hardy}},\ }%
  \bibfield{journal}{%
  \bibinfo {journal} {Phys. Rev. B}\ }%
  \textbf{\bibinfo {volume} {73}},\ \bibinfo {pages} {180505(R)} (\bibinfo
  {year} {2006})%
  \bibAnnoteFile{NoStop}{liang06}%
\bibitem{hosseini04}%
  \BibitemOpen
  \bibfield{author}{%
  \bibinfo {author} {\bibfnamefont{A.}~\bibnamefont{Hosseini}}, \bibinfo
  {author} {\bibfnamefont{D.~M.}\ \bibnamefont{Broun}}, \bibinfo {author}
  {\bibfnamefont{D.~E.}\ \bibnamefont{Sheehy}}, \bibinfo {author}
  {\bibfnamefont{T.~P.}\ \bibnamefont{Davis}}, \bibinfo {author}
  {\bibfnamefont{M.}~\bibnamefont{Franz}}, \bibinfo {author}
  {\bibfnamefont{W.~N.}\ \bibnamefont{Hardy}}, \bibinfo {author}
  {\bibfnamefont{R.}~\bibnamefont{Liang}},\ and\ \bibinfo {author}
  {\bibfnamefont{D.~A.}\ \bibnamefont{Bonn}},\ }%
  \bibfield{journal}{%
  \bibinfo {journal} {Phys. Rev. Lett.}\ }%
  \textbf{\bibinfo {volume} {93}},\ \bibinfo {pages} {107003} (\bibinfo {year}
  {2004})%
  \bibAnnoteFile{NoStop}{hosseini04}%
\bibitem{huttema}%
  \BibitemOpen
  \bibinfo {note} {W. Huttema (private communication).}%
  \bibAnnoteFile{Stop}{huttema}%
\bibitem{strongin70}%
  \BibitemOpen
  \bibfield{author}{%
  \bibinfo {author} {\bibfnamefont{M.}~\bibnamefont{Strongin}}, \bibinfo
  {author} {\bibfnamefont{R.~S.}\ \bibnamefont{Thompson}}, \bibinfo {author}
  {\bibfnamefont{O.~F.}\ \bibnamefont{Kammerer}},\ and\ \bibinfo {author}
  {\bibfnamefont{J.~E.}\ \bibnamefont{Crow}},\ }%
  \bibfield{journal}{%
  \bibinfo {journal} {Phys. Rev. B}\ }%
  \textbf{\bibinfo {volume} {1}},\ \bibinfo {pages} {1078} (\bibinfo {year}
  {1970}),\ \bibinfo {note} {(the superconductor-insulator transition is shown
  in Fig. 11)}%
  \bibAnnoteFile{NoStop}{strongin70}%
\bibitem{ferrell58}%
  \BibitemOpen
  \bibfield{author}{%
  \bibinfo {author} {\bibfnamefont{R.~A.}\ \bibnamefont{Ferrell}}\ and\
  \bibinfo {author} {\bibfnamefont{R.~E.}\ \bibnamefont{Glover{, }III}},\ }%
  \bibfield{journal}{%
  \bibinfo {journal} {Phys. Rev.}\ }%
  \textbf{\bibinfo {volume} {109}},\ \bibinfo {pages} {1398} (\bibinfo {year}
  {1958})%
  \bibAnnoteFile{NoStop}{ferrell58}%
\bibitem{basov05}%
  \BibitemOpen
  \bibfield{author}{%
  \bibinfo {author} {\bibfnamefont{D.~N.}\ \bibnamefont{Basov}}\ and\ \bibinfo
  {author} {\bibfnamefont{T.}~\bibnamefont{Timusk}},\ }%
  \bibfield{journal}{%
  \bibinfo {journal} {Rev. Mod. Phys.}\ }%
  \textbf{\bibinfo {volume} {77}},\ \bibinfo {pages} {721} (\bibinfo {year}
  {2005})%
  \bibAnnoteFile{NoStop}{basov05}%
\bibitem{kamaras90}%
  \BibitemOpen
  \bibfield{author}{%
  \bibinfo {author} {\bibfnamefont{K.}~\bibnamefont{Kamar{\'a}s}}, \bibinfo
  {author} {\bibfnamefont{S.~L.}\ \bibnamefont{Herr}}, \bibinfo {author}
  {\bibfnamefont{C.~D.}\ \bibnamefont{Porter}}, \bibinfo {author}
  {\bibfnamefont{N.}~\bibnamefont{Tache}}, \bibinfo {author}
  {\bibfnamefont{D.~B.}\ \bibnamefont{Tanner}}, \bibinfo {author}
  {\bibfnamefont{S.}~\bibnamefont{Etemad}}, \bibinfo {author}
  {\bibfnamefont{T.}~\bibnamefont{Venkatesan}}, \bibinfo {author}
  {\bibfnamefont{E.}~\bibnamefont{Chase}}, \bibinfo {author}
  {\bibfnamefont{A.}~\bibnamefont{Inam}}, \bibinfo {author}
  {\bibfnamefont{X.~D.}\ \bibnamefont{Wu}}, \bibinfo {author}
  {\bibfnamefont{M.~S.}\ \bibnamefont{Hegde}},\ and\ \bibinfo {author}
  {\bibfnamefont{B.}~\bibnamefont{Dutta}},\ }%
  \bibfield{journal}{%
  \bibinfo {journal} {Phys. Rev. Lett.}\ }%
  \textbf{\bibinfo {volume} {64}},\ \bibinfo {pages} {1692} (\bibinfo {year}
  {1990})%
  \bibAnnoteFile{NoStop}{kamaras90}%
\bibitem{hardy93}%
  \BibitemOpen
  \bibfield{author}{%
  \bibinfo {author} {\bibfnamefont{W.~N.}\ \bibnamefont{Hardy}}, \bibinfo
  {author} {\bibfnamefont{D.~A.}\ \bibnamefont{Bonn}}, \bibinfo {author}
  {\bibfnamefont{D.~C.}\ \bibnamefont{Morgan}}, \bibinfo {author}
  {\bibfnamefont{R.}~\bibnamefont{Liang}},\ and\ \bibinfo {author}
  {\bibfnamefont{K.}~\bibnamefont{Zhang}},\ }%
  \bibfield{journal}{%
  \bibinfo {journal} {Phys. Rev. Lett.}\ }%
  \textbf{\bibinfo {volume} {70}},\ \bibinfo {pages} {3999} (\bibinfo {year}
  {1993})%
  \bibAnnoteFile{NoStop}{hardy93}%
\bibitem{harlingen95}%
  \BibitemOpen
  \bibfield{author}{%
  \bibinfo {author} {\bibfnamefont{D.~J.}\ \bibnamefont{Van~Harlingen}},\ }%
  \bibfield{journal}{%
  \bibinfo {journal} {Rev. Mod. Phys.}\ }%
  \textbf{\bibinfo {volume} {67}},\ \bibinfo {pages} {515} (\bibinfo {year}
  {1995})%
  \bibAnnoteFile{NoStop}{harlingen95}%
\bibitem{basov94a}%
  \BibitemOpen
  \bibfield{author}{%
  \bibinfo {author} {\bibfnamefont{D.~N.}\ \bibnamefont{Basov}}, \bibinfo
  {author} {\bibfnamefont{T.}~\bibnamefont{Timusk}}, \bibinfo {author}
  {\bibfnamefont{B.}~\bibnamefont{Dabrowski}},\ and\ \bibinfo {author}
  {\bibfnamefont{J.~D.}\ \bibnamefont{Jorgensen}},\ }%
  \bibfield{journal}{%
  \bibinfo {journal} {Phys. Rev. B}\ }%
  \textbf{\bibinfo {volume} {50}},\ \bibinfo {pages} {3511} (\bibinfo {year}
  {1994})%
  \bibAnnoteFile{NoStop}{basov94a}%
\bibitem{shibauchi94}%
  \BibitemOpen
  \bibfield{author}{%
  \bibinfo {author} {\bibfnamefont{T.}~\bibnamefont{Shibauchi}}, \bibinfo
  {author} {\bibfnamefont{H.}~\bibnamefont{Kitano}}, \bibinfo {author}
  {\bibfnamefont{K.}~\bibnamefont{Uchinokura}}, \bibinfo {author}
  {\bibfnamefont{A.}~\bibnamefont{Maeda}}, \bibinfo {author}
  {\bibfnamefont{T.}~\bibnamefont{Kimura}},\ and\ \bibinfo {author}
  {\bibfnamefont{K.}~\bibnamefont{Kishio}},\ }%
  \bibfield{journal}{%
  \bibinfo {journal} {Phys. Rev. Lett.}\ }%
  \textbf{\bibinfo {volume} {72}},\ \bibinfo {pages} {2263} (\bibinfo {year}
  {1994})%
  \bibAnnoteFile{NoStop}{shibauchi94}%
\bibitem{carr89}%
  \BibitemOpen
  \bibfield{author}{%
  \bibinfo {author} {\bibfnamefont{G.~L.}\ \bibnamefont{Carr}}\ and\ \bibinfo
  {author} {\bibfnamefont{D.~B.}\ \bibnamefont{Tanner}},\ }%
  \bibfield{journal}{%
  \bibinfo {journal} {Phys. Rev. Lett.}\ }%
  \textbf{\bibinfo {volume} {62}},\ \bibinfo {pages} {2763} (\bibinfo {year}
  {1989})%
  \bibAnnoteFile{NoStop}{carr89}%
\bibitem{lang02}%
  \BibitemOpen
  \bibfield{author}{%
  \bibinfo {author} {\bibfnamefont{K.~M.}\ \bibnamefont{Lang}}, \bibinfo
  {author} {\bibfnamefont{V.}~\bibnamefont{Madhavan}}, \bibinfo {author}
  {\bibfnamefont{J.~E.}\ \bibnamefont{Hoffman}}, \bibinfo {author}
  {\bibfnamefont{E.~W.}\ \bibnamefont{Hudson}}, \bibinfo {author}
  {\bibfnamefont{H.}~\bibnamefont{Eisaki}}, \bibinfo {author}
  {\bibfnamefont{S.}~\bibnamefont{Uchida}},\ and\ \bibinfo {author}
  {\bibfnamefont{J.~C.}\ \bibnamefont{Davis}},\ }%
  \bibfield{journal}{%
  \bibinfo {journal} {Nature (London)}\ }%
  \textbf{\bibinfo {volume} {415}},\ \bibinfo {pages} {412} (\bibinfo {year}
  {2002})%
  \bibAnnoteFile{NoStop}{lang02}%
\bibitem{mcelroy05a}%
  \BibitemOpen
  \bibfield{author}{%
  \bibinfo {author} {\bibfnamefont{K.}~\bibnamefont{McElroy}}, \bibinfo
  {author} {\bibfnamefont{J.}~\bibnamefont{Lee}}, \bibinfo {author}
  {\bibfnamefont{J.~A.}\ \bibnamefont{Slezak}}, \bibinfo {author}
  {\bibfnamefont{D.-H.}\ \bibnamefont{Lee}}, \bibinfo {author}
  {\bibfnamefont{H.}~\bibnamefont{Eisaki}}, \bibinfo {author}
  {\bibfnamefont{S.}~\bibnamefont{Uchida}},\ and\ \bibinfo {author}
  {\bibfnamefont{J.~C.}\ \bibnamefont{Davis}},\ }%
  \bibfield{journal}{%
  \bibinfo {journal} {Science}\ }%
  \textbf{\bibinfo {volume} {309}},\ \bibinfo {pages} {1048} (\bibinfo {year}
  {2005})%
  \bibAnnoteFile{NoStop}{mcelroy05a}%
\bibitem{dagotto05}%
  \BibitemOpen
  \bibfield{author}{%
  \bibinfo {author} {\bibfnamefont{E.}~\bibnamefont{Dagotto}},\ }%
  \bibfield{journal}{%
  \bibinfo {journal} {Science}\ }%
  \textbf{\bibinfo {volume} {309}},\ \bibinfo {pages} {257} (\bibinfo {year}
  {2005})%
  \bibAnnoteFile{NoStop}{dagotto05}%
\bibitem{zaanen89}%
  \BibitemOpen
  \bibfield{author}{%
  \bibinfo {author} {\bibfnamefont{J.}~\bibnamefont{Zaanen}}\ and\ \bibinfo
  {author} {\bibfnamefont{O.}~\bibnamefont{Gunnarsson}},\ }%
  \bibfield{journal}{%
  \bibinfo {journal} {Phys. Rev. B}\ }%
  \textbf{\bibinfo {volume} {40}},\ \bibinfo {pages} {7391(R)} (\bibinfo {year}
  {1989})%
  \bibAnnoteFile{NoStop}{zaanen89}%
\bibitem{tranquada95}%
  \BibitemOpen
  \bibfield{author}{%
  \bibinfo {author} {\bibfnamefont{J.~M.}\ \bibnamefont{Tranquada}}, \bibinfo
  {author} {\bibfnamefont{B.~J.}\ \bibnamefont{Sternlieb}}, \bibinfo {author}
  {\bibfnamefont{J.~D.}\ \bibnamefont{Axe}}, \bibinfo {author}
  {\bibfnamefont{Y.}~\bibnamefont{Nakamura}},\ and\ \bibinfo {author}
  {\bibfnamefont{S.}~\bibnamefont{Uchida}},\ }%
  \bibfield{journal}{%
  \bibinfo {journal} {Nature (London)}\ }%
  \textbf{\bibinfo {volume} {375}},\ \bibinfo {pages} {561} (\bibinfo {year}
  {1995})%
  \bibAnnoteFile{NoStop}{tranquada95}%
\bibitem{emery93}%
  \BibitemOpen
  \bibfield{author}{%
  \bibinfo {author} {\bibfnamefont{V.~J.}\ \bibnamefont{Emery}}\ and\ \bibinfo
  {author} {\bibfnamefont{S.~A.}\ \bibnamefont{Kivelson}},\ }%
  \bibfield{journal}{%
  \bibinfo {journal} {Physica C}\ }%
  \textbf{\bibinfo {volume} {209}},\ \bibinfo {pages} {597} (\bibinfo {year}
  {1993})%
  \bibAnnoteFile{NoStop}{emery93}%
\bibitem{simanek}%
  \BibitemOpen
  \bibfield{author}{%
  \bibinfo {author} {\bibfnamefont{E.}~\bibnamefont{{\v{S}}im{\'{a}}nek}},\ }%
  \emph{\bibinfo {title} {Inhomogeneous Superconductors}}\ (\bibinfo
  {publisher} {Oxford University Press},\ \bibinfo {address} {New York},\
  \bibinfo {year} {1994})%
  \bibAnnoteFile{NoStop}{simanek}%
\bibitem{imry08}%
  \BibitemOpen
  \bibfield{author}{%
  \bibinfo {author} {\bibfnamefont{Y.}~\bibnamefont{Imry}}, \bibinfo {author}
  {\bibfnamefont{M.}~\bibnamefont{Strongin}},\ and\ \bibinfo {author}
  {\bibfnamefont{C.~C.}\ \bibnamefont{Homes}},\ }%
  \bibfield{journal}{%
  \bibinfo {journal} {Physica C}\ }%
  \textbf{\bibinfo {volume} {468}},\ \bibinfo {pages} {288} (\bibinfo {year}
  {2008})%
  \bibAnnoteFile{NoStop}{imry08}%
\bibitem{imry}%
  \BibitemOpen
  \bibinfo {note} {Y. Imry (private communication).}%
  \bibAnnoteFile{Stop}{imry}%
\end{thebibliography}%

\end{document}